\documentclass[conference]{IEEEtran}
\IEEEoverridecommandlockouts
\usepackage{cite}
\usepackage{amsmath,amssymb,amsfonts}
\usepackage{algorithmic}
\usepackage{graphicx}
\usepackage{textcomp}
\usepackage{xcolor}

\usepackage{graphics} 
\usepackage{epsfig} 
\usepackage{mathptmx} 
\usepackage{times} 
\usepackage{amsmath} 
\usepackage{amssymb}  
\usepackage{subcaption}
\usepackage{multirow}
\usepackage{graphicx}
\usepackage{amsmath}
\usepackage{threeparttable,booktabs}

\def\BibTeX{{\rm B\kern-.05em{\sc i\kern-.025em b}\kern-.08em
    T\kern-.1667em\lower.7ex\hbox{E}\kern-.125emX}}
    
\usepackage{graphicx}

\DeclareRobustCommand*{\IEEEauthorrefmark}[1]{%
  \raisebox{0pt}[0pt][0pt]{\textsuperscript{\footnotesize #1}}%
}

\begin{document}

\author{
    \IEEEauthorblockN{Xinwei Ju\IEEEauthorrefmark{†}, Frank Po Wen Lo\IEEEauthorrefmark{†,*}, Jianing Qiu, Peilun Shi, Jiachuan Peng, Benny Lo}
    \IEEEauthorblockA{Hamlyn Centre, Imperial College London, UK.
    \\\{x.ju21, po.lo15, jianing.qiu17, p.shi21, j.peng21, benny.lo\}@imperial.ac.uk}
   
    }

\title{MenuAI: Restaurant Food Recommendation System via a Transformer-based Deep Learning Model
{}
\thanks{† refers to equal contributions; $*$ refers to the corresponding author\\
This project is supported by the Innovative Passive Dietary Monitoring Project funded by Bill \& Melinda Gates Foundation (Opportunity ID: INV-006713).}
}

\maketitle

\begin{abstract}

Food recommendation system has proven as an effective technology to provide guidance on dietary choices, and this is especially important for patients suffering from chronic diseases. Unlike other multimedia recommendations, such as books and movies, food recommendation task is highly relied on the context at the moment, since users' food preference can be highly dynamic over time. For example, individuals tend to eat more calories earlier in the day and eat a little less at dinner. However, there are still limited research works trying to incorporate both current context and nutritional knowledge for food recommendation. Thus, a novel restaurant food recommendation system is proposed in this paper to recommend food dishes to users according to their special nutritional needs. Our proposed system utilises Optical Character Recognition (OCR) technology and a transformer-based deep learning model, Learning to Rank (LTR) model, to conduct food recommendation. Given a single RGB image of the menu, the system is then able to rank the food dishes in terms of the input search key (e.g., calorie, protein level). Due to the property of the transformer, our system can also rank unseen food dishes. Comprehensive experiments are conducted to validate our methods on a self-constructed menu dataset, known as MenuRank dataset. The promising results, with accuracy ranging from 77.2\% to 99.5\%, have demonstrated the great potential of LTR model in addressing food recommendation problems.

\end{abstract}
\begin{figure}[t]
\centering
\includegraphics[width=\columnwidth]{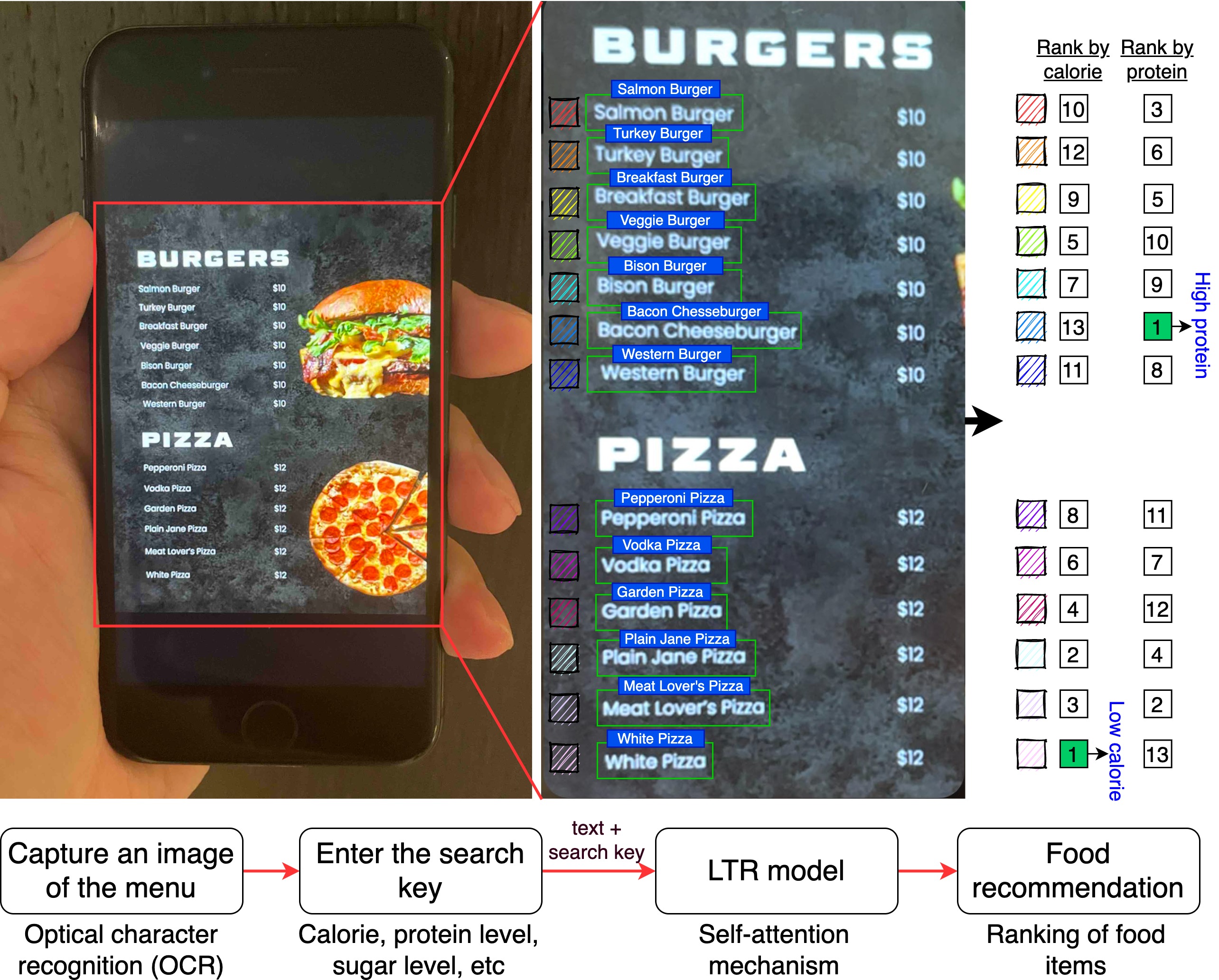}
\caption{\textbf{Pipeline of the proposed food recommendation system - MenuAI.} Given a single image of the menu, our proposed system is able to rank the dishes in terms of the search key (e.g., calorie, protein level, etc). Note that the system can also infer the ranking of unseen food items.}
\label{fig:1}
\vspace{-5pt}
\end{figure}

\begin{figure*}[t]
\centering
\includegraphics[width=0.95\textwidth]{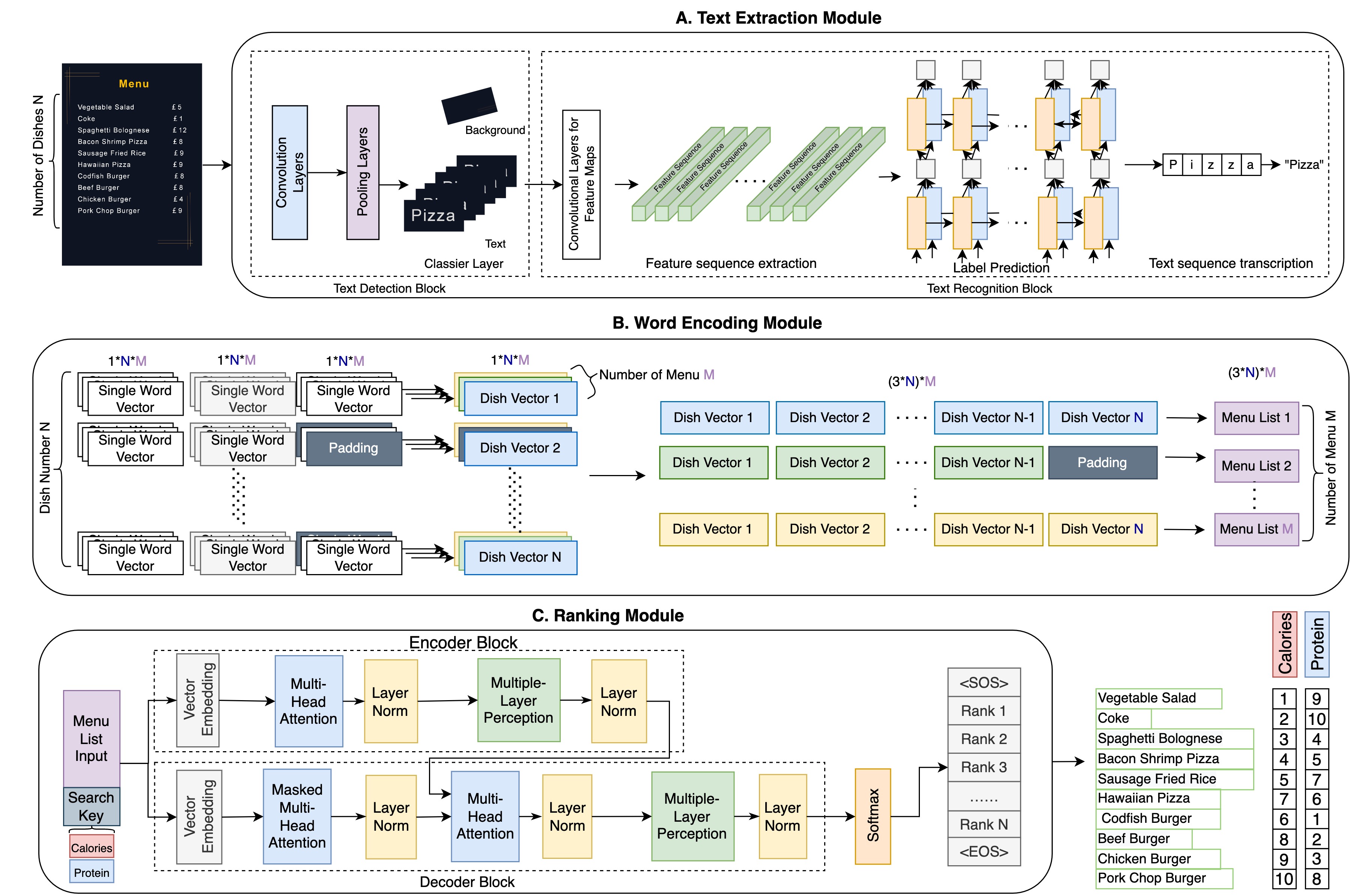}
\caption{\textbf{The network architecture of our proposed food recommendation system - MenuAI.} The system can be divided into three main modules: (A) Text extraction module (B) Word encoding module (C) Ranking module}
\label{fig:2}
\vspace{-5pt}
\end{figure*}

\vspace{-1mm}
\section{Introduction}
 According to the survey \cite{jones_tong_monsivais_2018}, the most important behavioural determinants of health are dietary choices, and unhealthy diet could further endanger the life of individuals suffering from chronic diseases, such as diabetes, cardiovascular disease and malignancies, etc \cite{lo2019point2volume}. The Global Burden of Disease (GBD) reported that inappropriate diets might have caused tens of millions of preventable premature deaths each year \cite{james2018global}. From these findings, we notice that it is important for patients with chronic diseases to properly control their eating behaviour and select appropriate food dishes for eating. Regarding dietary choices, it is easy to determine what to eat at home since we are familiar with the ingredients that we eat on a daily basis. However, decisions cannot be made effectively without in-depth understanding of the food items on the menu when we eat out at restaurants. Thus, food recommendation systems play an important role in providing guidance for individuals to select the most appropriate food dishes. 
 
 Recent advances in artificial intelligence and matrix completion technique have enhanced the efficiency of recommendation in various aspects (e.g., music, movies and books), but there are limited research works in the food domain. Nowadays, users' preference and nutritional needs are two main considerations for a food recommendation system. For example, Nag \textit{et~al.} \cite{nag2017live} developed a personalised food recommendation system by fusing contextually-aware (i.e., barometer and pedometer) and personalised data in order to calculate daily values by a pre-defined equation. Then the calculated daily values are used to rank the dishes for food recommendation. Ribeiro \textit{et~al.} \cite{ribeiro2017souschef} proposed to consider more factors in the recommendation system which include nutrition, users' food preference and the budget. 

Despite the fact that these methods have shown promises in food recommendation, several key challenge are still unresolved. (1) Knowledge Graph (KG) has been extensively used in existing multimedia recommendation system (e.g., movies and books), in which it obtains and integrates data into an ontology and makes use of reasoners to infer new knowledge; however there are limited research works focusing on food-oriented KG \cite{8930090}. Thus, unseen food recommendation (i.e., food dishes not in the nutritional database) is still challenging because of the lack of associations between food dishes. (2) Existing food recommendation systems always assume that users' food preference is static, but it should be dynamic over time \cite{delarue2004dynamics}.

To address the aforementioned problems, we proposed a transformer-based deep learning model, known as Learning to Rank (LTR) model, to conduct food recommendation via Optical Character Recognition (OCR), self-attention mechanism and pair-wise comparison. The proposed LTR model can be trained end-to-end to learn the ranking of the food dishes with regards to the input search key (i.e., calorie, protein level, sugar level, etc). Furthermore, due to the property of the transformer-based network architecture, our system can also rank unseen food dishes. The problem can be formulated as follows: Given an input matrix $\textbf{X}\in\mathbb{R}^{N\times M }$, where \textbf{X} is composed of a list of input vectors $[x_{1}, x_{2},...,x_{M}]$. The vector $x_i$ refers to the encoded text (i.e., dimension $N$) representing a single food dish on the menu. Note that the number of input vectors (i.e., food dishes) can be of varying length. The network is then trained to minimise the loss function between the estimated ranking order and the ground truth label with regards to the search key. The main contributions of this paper can be summarised as follows: (1) A novel transformer-based deep learning model, LTR model, is proposed to carry out food recommendation, in which OCR and self-attention mechanism are first combined and introduced to rank the food dishes. (2) The ranking orders of food dishes can be obtained with regards to specific nutritional needs by using corresponding search keys. (3) The system has reasoning ability which can predict the rank of unseen food dishes.

\section{Methodology}

The pipeline of our proposed food recommendation system, as shown in Fig.\ref{fig:1}, can be divided into the following steps. (1) A mobile phone is required to capture a single RGB image of the menu. (2) Menu digitisation is conducted through OCR, and the text information is extracted from the image. (3) Each of the extracted word is encoded into a single word vector and then several single word vectors are grouped into a single dish vector to represent a food dish. (4) The dish vectors are padded and packed into an input matrix and the input matrix is then fed into the ranking module, the LTR model, for training a scoring function for food recommendation. Note that our proposed LTR model can be used to handle the menu with varying number of food dishes.

\subsection{Text extraction}

Our OCR technique (i.e., menu digitisation) is implemented based on \cite{ocrpytorch}. This technique is the process of analysing and recognising photographs to extract text information and convert them into digital form. The task of text extraction can mainly be divided into two parts - text detection via Connectionist Text Proposal Network (CTPN) \cite{tian2016detecting} and text recognition via Convolutional Recurrent Neural Network (CRNN) \cite{7801919}. CTPN uses an anchored regression mechanism to accurately localise text lines in natural image and predict the text position's confidence score. For CRNN, it is consisted of three parts. The convolution layers are used to extract feature, and the recurrent layers are used for predicting tags (i.e., the orange and blue blocks, shown in Fig. \ref{fig:2}A, represent
the forward and reverse RNN respectively). Last, the transcription layer is to transform text sequences.

\subsection{Word encoding}

The extracted text information cannot be processed directly by the LTR model. Thus, word encoding is conduced to encode the text and convert them into feature vectors. Despite one-hot encoding show promises in traditional Natural Language Processing (NLP), there are several drawbacks such as dimension disaster and high memory load. Thus, we consider using index-based word encoding, which involves creating a dictionary and assigning a unique index to all conceivable words so that we can uniquely identify them using the index (i.e. also known as single word vector). The single word vector can then be grouped into a dish vector, as shown in Fig. \ref{fig:2}B. Note that we have standardised food dishes to make sure they are consisted of maximum three words. To ensure the input matrix which is fed into the LTR module has consistent dimension, dishes with fewer than three words will have their vectors padded with zeros. Note that BERT encoder will be applied to facilitate the text encoding in our future work.

\begin{figure}[t]
\centering
\includegraphics[width=\columnwidth]{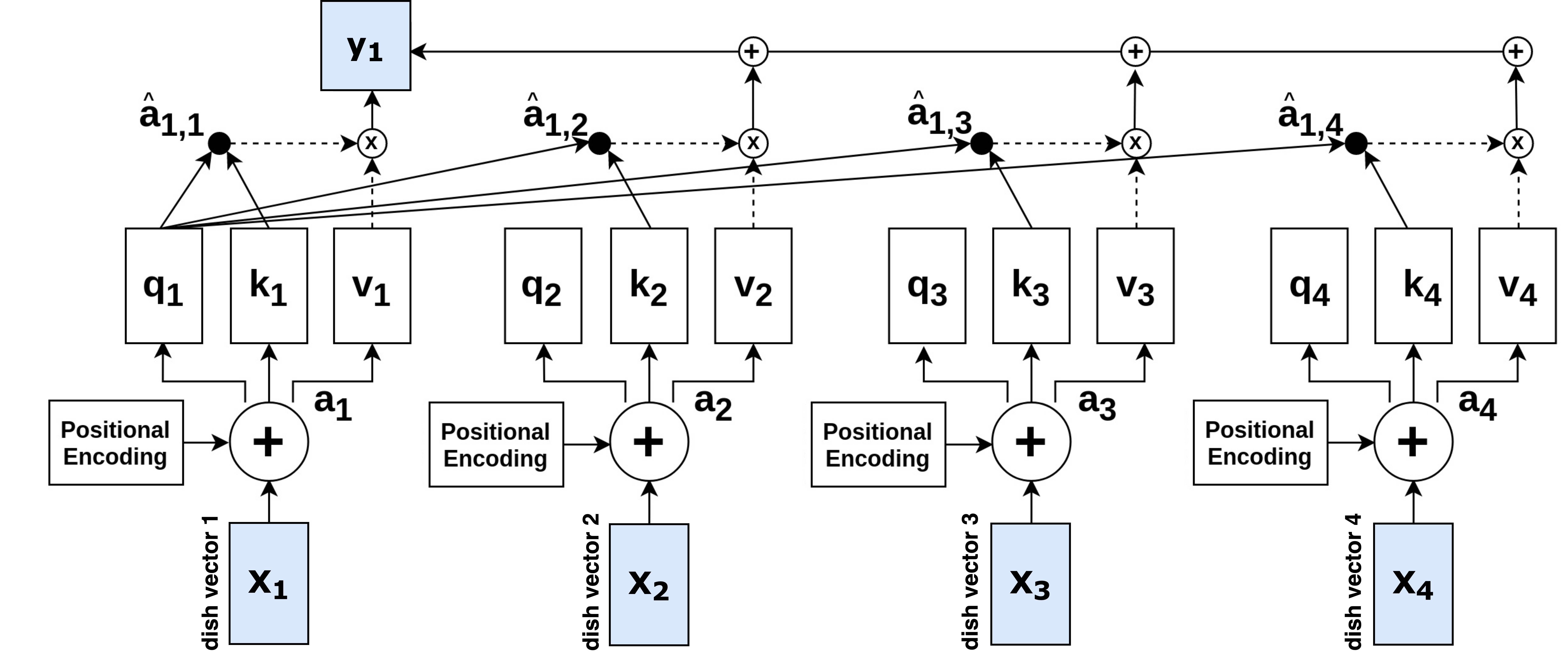}
\caption{\textbf{Self-attention mechanism.} The ranking of each food dish $y_i$ is determined with the context of all other food dishes presented on the menu [$x_1$, $x_2$,...,$x_n$]. Note that the LTR network architecture is implemented based on our previous work  \cite{lo2021deep3dranker}. 
}
\label{fig:3}
\vspace{-7pt}
\end{figure}

\subsection{Ranking module - LTR model}

Given the dish feature vectors, they will be padded and packed into an input matrix. Then both the input matrix and the search key will be fed into the ranking module, the LTR model, in which the optimal order with regards to the specific key can be inferred, as shown in Fig. \ref{fig:2}C. In our work, the LTR model utilises the technique of self-attention mechanism implemented based on our previous work \cite{lo2021deep3dranker}. The reason of using such an approach is that we require each dish feature vector to carry our pair-wise comparison with other dishes before prediction while the self-attention mechanism can help facilitate the context-aware learning. As shown in Fig. \ref{fig:3}, the rationale of the self-attention mechanism can be presented as mapping three vectors (i.e., \textit{value} ($v_{i}$), \textit{key} ($k_{i}$) and \textit{query} ($q_{i}$)) to higher level representations by taking a weighted sum of the \textit{values} over all food dishes. The weight assigned to each \textit{value} is relied on the relevance of each \textit{key} item to the \textit{query} item. The output $y_{1}$ can then be calculated as follows:  

\begin{equation}
y_{1} = \sum_{i}^{N}\hat{a}_{1,i}v_{i}
\label{weighted}
\end{equation}
where $\hat{a}_{1,i}$ is the weight corresponding to the value $v_{i}$. In Eq. (\ref{weighted}), we can notice that each output $y_{i}$ is calculated after considering the whole sequence $v_i$ as well as $x_{i}$ intuitively, making the attention mechanism effective for ranking \cite{lo2021deep3dranker}.

\section{Experimental results and discussions}

\subsection{Data preparation}

To examine the performance of the LTR model for restaurant food recommendation, a large-scale menu dataset with ranking annotations with regards to specific search key is needed. A self-constructed dataset, known as MenuRank, with 5625 menus is generated. In each menu, food dishes are randomly included and the number of food dishes on each menu is also randomly ranging from 7 to 15 (i.e., Note that this paper aims to conduct a pilot study to evaluate the effectiveness of LTR model in food recommendations. Only 38 different food servings are used in total). The ranking order of food dishes in each menu is annotated manually based on a reputable nutritional dataset \cite{boohee}. Unless otherwise stated, we train our models with train:test sets of a 80:20 split (i.e., 4500:1125).

\subsection{Ranking results on MenuRank dataset}
To evaluate the effectiveness of the proposed food recommendation system on ranking food dishes by calories, we first train the system using the MenuRank dataset with single search key (i.e., menus with calories annotated). A rank-aware metric, Normalised Discounted Cumulative Gain (NDCG), is chosen to evaluate the model (i.e., NDCG has been extensively applied to evaluate the ranking performance of recommendation system). Apart from NDCG, Cross Entropy Loss (CEL) and Accuracy (ACC) are also calculated in the quantitative results. In Table \ref{tab1}, we first evaluate our models on menus with seen dishes (i.e., seen dishes refer to the dishes which include in those 38 food servings but with unseen combinations presented on the menu), the result is promising which can achieve up to 99.5\% in ACC. This preliminary result proves that the LTR model is effective in ranking seen food dishes with unseen combinations on the menus. Furthermore, to evaluate the reasoning capability of the LTR model, we further construct two test dataset with only 50\% seen food dishes and 10\% seen food dishes and the LTR model is evaluated on these dataset (i.e., each of them has 1125 menus as well). From Table \ref{tab1}, it shows that the proposed LTR model achieves satisfactory performance on both dataset, where the testing ACC are up to 86.5\% and 71.2\% respectively. We believe these findings could undoubtedly open opportunities for enhancing the performance of existing food recommendation systems. 

\begin{figure}[t]
\centering
\includegraphics[width=0.85\columnwidth]{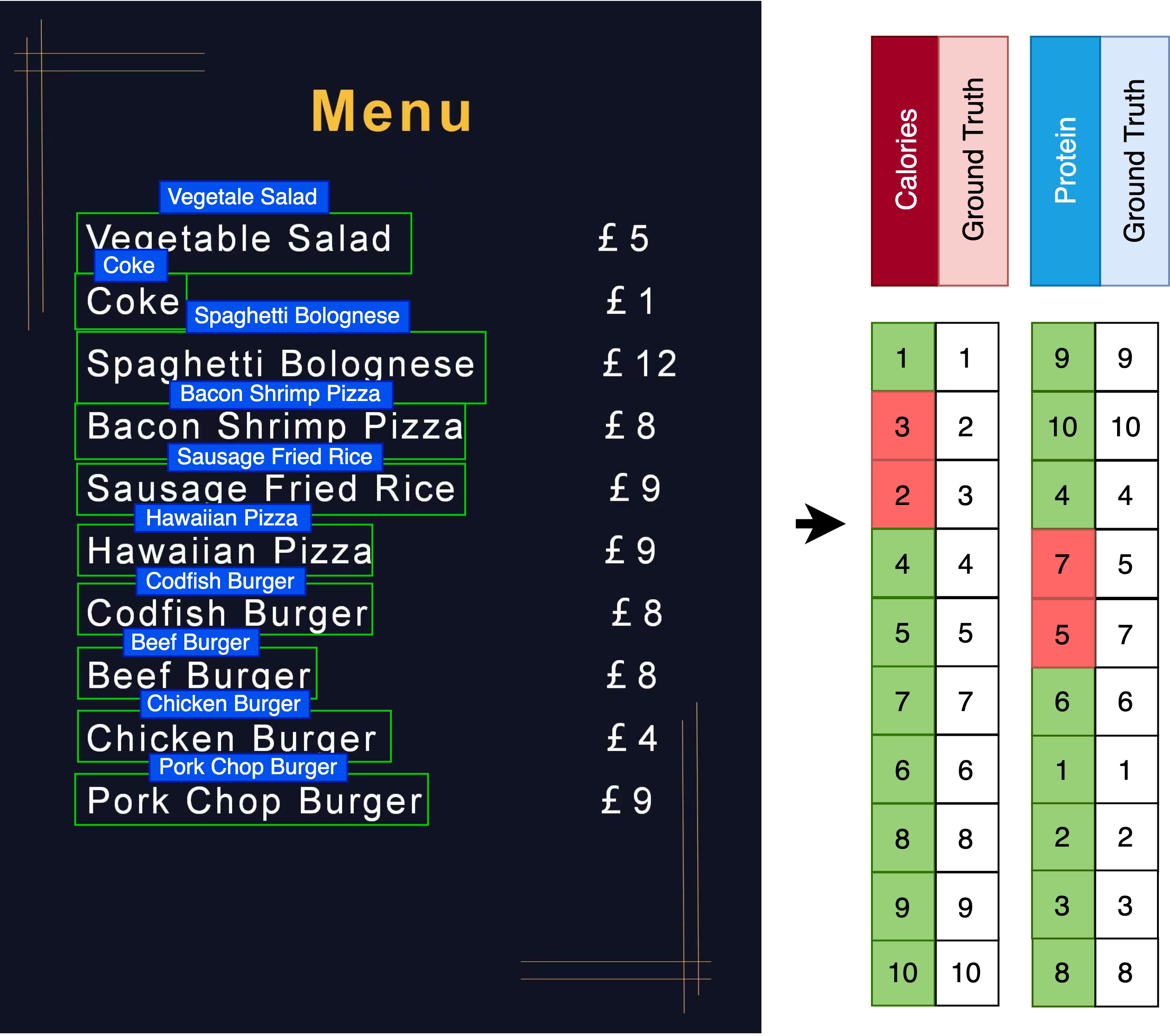}
\caption{\textbf{Performance of the proposed recommendation system.} Two different search keys (i.e., calories and protein level) are used to rank food dishes.}
\label{fig:4}
\vspace{-5pt}
\end{figure}

\subsection{Performance of the system with multiple search keys}
To further ease the training process of our system, we also train our LTR model using MenuRank dataset with multiple search keys (i.e., not just using calories ranking as ground truth in the training dataset). The advantage of using such an approach is that only a single trained model is required to conduct food recommendation with different search keys. In Table \ref{tab2},  we evaluate our models on menus with seen dishes, 50\% seen dishes and 10\% seen dishes. We noticed that the experimental results are still promising with a slight drop in ACC (i.e., 96.1\%, 71.7\% and 65.2\% in ACC respectively). But the performance can be gradually improved when the training dataset is scaled up. More future works can be done to further improve the performance of LTR models trained with multiple search keys. Fig. \ref{fig:4} gives an example of our proposed recommendation systems to show how it works under multiple search keys.

\begin{table}[t]
\caption{Quantitative results on food recommendation (with a single search key)}
\label{tab1}
\resizebox{\columnwidth}{!}{%
\begin{tabular}{c|c|c|c}
\hline
     & \textbf{100\% Seen Dishes} & \textbf{50\% Seen Dishes} & \textbf{10\% Seen Dishes} \\ \hline
NDCG↑ & 0.986                      & 0.865                     & 0.712                     \\
CEL↓  & 0.001                      & 0.065                     & 0.268                     \\
ACC↑  & 0.995                      & 0.857                     & 0.772                     \\ \hline
\end{tabular}%
}

\begin{tablenotes}
\item[] \scriptsize
NDCG: Normalized Discounted Cumulative Gain; CEL: Cross Entropy Loss; ACC: Accuracy; Note that $\uparrow$ the larger is better; $\downarrow$ the smaller is better
\end{tablenotes} 
\vspace{-10pt}
\end{table}

\begin{table}[t]
\caption{Quantitative results on food recommendation (with multiple search keys)}
\label{tab2}
\resizebox{\columnwidth}{!}{%
\begin{tabular}{c|c|c|c}
\hline
     & \textbf{100\% Seen Dishes} & \textbf{50\% Seen Dishes} & \textbf{10\% Seen Dishes} \\ \hline
NDCG↑ & 0.895                      & 0.798                     & 0.685                     \\
CEL↓  & 0.013                      & 0.223                     & 0.295                     \\
ACC↑  & 0.961                      & 0.717                     & 0.652                     \\ \hline
\end{tabular}%
}
\vspace{-10pt}

\end{table}

\section{Conclusion}

In this paper, we tackle several new problems in food recommendation and implement a novel transformer-based deep learning model to provide guidance on dietary choices in restaurants via OCR technology and self-attention mechanism. Unlike the existing food recommendation systems, our proposed system allows users to rank the food dishes with regards to their specific nutritional needs (e.g., calories, protein level, etc) by using the search keys function. Furthermore, due to the property of the transformer-based network architecture, our system can also rank unseen food dishes. The proposed system have demonstrated the effectiveness in ranking food dishes with any combination and unseen food dishes through the extensive experiments on our self-constructed dataset - MenuRank dataset. We believe that this research work will open up new open opportunities for enhancing the performance of existing food recommendation systems.

\bibliographystyle{IEEEtran}
\bibliography{ref}

\end{document}